# Roles of Defects and Sb-doping in the Thermoelectric Properties of Full-Heusler $Fe_2TiSn$

*Ilaria Pallecchi [a,‡], Daniel I. Bilc [*,b,‡], Marcella Pani [c,a], Fabio Ricci [d], Sébastien Lemal [d], Philippe Ghosez [d], Daniele Marré [e,a]*

[a] CNR-SPIN, Dipartimento di Fisica, Via Dodecaneso 33, 16146, Genova, Italy

[b] Faculty of Physics, Babeş-Bolyai University, 1 Kogalniceanu, RO-400084 Cluj-Napoca, România

[c] Dipartimento di Chimica e Chimica Industriale, Università di Genova, Via Dodecaneso 31, I-16146 Genova, Italy

[d] Theoretical Materials Physics, Q-MAT, CESAM, Université de Liège (B5), B-4000 Liège, Belgium

[e] Dipartimento di Fisica, Università di Genova, Via Dodecaneso 31, I-16146 Genova, Italy

ABSTRACT. The potential of $Fe_2TiSn$ full-Heusler compounds for thermoelectric applications has been suggested theoretically, but not yet grounded experimentally, due to the difficulty of obtaining reproducible, homogeneous, phase pure and defect free samples. In this work, we study $Fe_2TiSn_{1-x}Sb_x$ polycrystals (x from 0 to 0.6), fabricated by high-frequency melting and long-time high-temperature annealing. We obtain fairly good phase purity, homogeneous microstructure and good matrix stoichiometry. Although intrinsic p-type transport behavior is dominant, n-type charge compensation by Sb doping is demonstrated. Calculations of formation energy of defects and electronic properties carried out in the density functional theory formalism reveal that charged iron vacancies $V_{Fe}^{2-}$ are the dominant defects responsible for the intrinsic p-type doping of $Fe_2TiSn$ in all types of growing conditions except Fe-rich. Additionally, Sb substitutions at Sn site give rise either to $Sb_{Sn}$, $Sb_{Sn}^{1+}$ which are responsible for n-type doping and magnetism ($Sb_{Sn}$) or to magnetic $Sb_{Sn}^{1-}$ which act as additional p-type dopants. Our experimental data highlight good thermoelectric properties close to room temperature, with Seebeck coefficients up to 56 µV/K in the x=0.2 sample and power factors up to $4.8 \times 10^{-4}$ W m$^{-1}$ K$^{-2}$ in the x=0.1 sample. Our calculations indicate the appearance of a pseudogap in Ti-rich conditions and large Sb doping, possibly improving further the thermoelectric properties.

## 1. Introduction

While half-Heusler compounds are known to be promising thermoelectric materials, due to their large Seebeck coefficient, high electrical conductivity and tunability of electronic and thermal properties by chemical substitution, [1] the potential of full-Heusler compounds for thermoelectric applications has been less intensively explored. First-principles calculations predicted that very large n-type power factors can be achieved in the family of $Fe_2YZ$ full-Heusler compounds (with $Y$ =Ti, Zr, Hf and $Z$ = Si, Ge, Sn) from the band engineering of highly directional Fe $e_g$ states. [2] Consistently with that, independent calculations reported that, thanks to flat bands as close as 0.04 eV to the Fermi level, [3] carrier dopings in the range $10^{20}$-$10^{21}$ cm$^{-3}$ in full-Heusler $Fe_2TiSn$ may yield enhanced thermoelectric properties and ZT values as large as 0.6 at room temperature. [4] Combined with their high thermal and chemical stabilities and the earth abondance of their constituent elements, these findings unveil the potential interest of full-Heusler compounds for thermoelectric applications.

Recent works have shown very large power factors $S^2\sigma$ of ~10.3x10$^{-3}$ W m$^{-1}$ K$^{-2}$ (resp. ~1.9x10$^{-3}$ W m$^{-1}$ K$^{-2}$) and thermoelectric figure of merit ZT of ~0.34 (resp. ~0.15), which were achieved through the engineering of band gap opening (resp. resonant states) in n-type $Fe_2V_{0.95}Ta_{0.05}Al_{0.9}Si_{0.1}$ alloys at 300 K [5,6,7] (resp. p-type $Fe_2VAl_{1.6}$ alloys at 400 K [8]) (ZT = $S^2\sigma T/\kappa$, S Seebeck coefficient, $\sigma$ electrical conductivity, $\kappa$ total thermal conductivity, and T - absolute temperature). These n-type $S^2\sigma$ values are even larger than those of the state-of-the-art thermoelectric materials, but the ZT values of full-Heusler compounds are smaller due to their larger lattice thermal conductivities. [9,10,11] Therefore, to improve their thermoelectric performance, it is important to better control the formation of defects and nanostructures with sharp boundary interfaces inside their matrix, similarly to what was achieved in other classes of materials. [9,11]

One of the main issues with Heusler compounds is their natural tendency to atomic disorder, namely vacancies, antisites, or swaps, [3,12,13,14] which make the optimized preparation protocol particularly critical. For instance, no univocal report on annealing temperatures and times can be found in literature. [3,15,16,17,18] More critically, such atomic disorder may affect transport properties.

[3,13] Antisite disorder was indeed demonstrated to be responsible for p-type conduction and semi-metallic behavior of undoped [3,19,20] and Sb-doped [20] $Fe_2TiSn$ annealed at different temperatures and for different times (at 700°C for 8 days, [20] at 800°C for 3 days, [19] and at 850°C for 7 days [3]), in contrast with the gapped band structure predicted by theory. Thermoelectric transport was measured in Sb doped $Fe_2TiSn$ above room temperature in samples annealed at 850°C [3] and below room temperature in samples annealed at 700°C. [20] In the former work, [3] it was found that p-type doping due to antisite disorder and n-type doping due to Sb substitution compensate each other, thus slightly decreasing the room temperature Seebeck coefficient with increasing Sb doping. In the latter work, [20] it was found that with doping up to 10%, the room temperature Seebeck coefficient and power factor $S^2\sigma$ are slightly increased, due to density of states features induced by native defects in the undoped and Sb doped samples, even if the measured p-type power factor values were much lower than the n-type power factor values predicted by theory. [2] It appears therefore that the preparation method and the annealing time and temperature protocol are crucial in determining the amount of antisite disorder and the microstructure, which further significantly affect the transport properties. Clarifying this further is therefore of key importance for optimizing the thermoelectric properties of full-Heusler compounds.

Here, we explore a wider range of Sb doping levels for $Fe_2TiSn$ as compared to our previous work, [20] and we adopt an optimized preparation method and annealing protocol. Specifically, we prepare $Fe_2TiSn_{1-x}Sb_x$ (x=0, 0.1, 0.2, 0.5 and 0.6) samples by high-frequency melting method and carry out annealing at 900°C for 13 days. We investigate the effect of Sb doping on the electric and thermoelectric transport properties. Our experimental measurements indicate that all the samples are p-type, as seen by the positive slope of the ordinary Hall effect and from the positive sign of the Seebeck effect, however, with increasing Sb substitution, the measured p-type carrier density decreases, highlighting that n-type doping compensation occurs. All the samples, except for the undoped one, show some evidence of magnetism, as seen from the anomalous contribution to the Hall effect and from negative magnetoresistance. Our calculations show that charged iron vacancies $V_{Fe}{}^{2-}$ have low formation energy, being associated to p-type intrinsic doping of $Fe_2TiSn$, while Sb doping at Sn site is responsible for n-type doping, as well as for magnetism together with the charged substitutional magnetic defects $Sb_{Sn}{}^{1-}$ in the substituted samples. A pseudogap formation in $Fe_2TiSn_{1-x}Sb_x$ is expected in Fe-rich conditions and at large Sb doping, which could yield a further enhancement of thermoelectric properties.

## 2. Methods

### 2.1. Experimental Methods

Polycrystalline samples of $Fe_2TiSn_{1-x}Sb_x$ (x = 0, 0.1, 0.2, 0.5 and 0.6) are prepared by high-frequency melting in an induction furnace in high-purity argon atmosphere and then annealed in vacuum at 900°C for 13 days. Higher doping x > 0.6 results in the formation of secondary phases, inhomogeneity, and elemental segregation, as also observed in ref. [19]

The microstructural homogeneity and stoichiometry of the samples is probed by scanning electron microscope - energy dispersive system (EDS) (Leica Cambridge S360, Oxford X-Max20 spectrometer, with software Aztec), averaging over multiple areas. Phase purity is analyzed by X-ray powder diffraction (XRPD).

Electrical resistivity (ρ), magnetoresistivity and Hall resistance ($R_H$) measurements by four-probe technique are performed in Physical Properties Measurement System (PPMS) by Quantum

Design at temperatures from room temperatures down to 10K and in magnetic fields up to 90000 Oe. The Seebeck (S) effect is measured with the PPMS Thermal Transport Option in continuous scanning mode with a 0.4 K/min cooling rate.

## 2.2. Computational Methods

The structural, electronic, and formation energy properties of different defects in $Fe_2TiSn_{1-x}Sb_x$ were studied within density functional theory (DFT) formalism using the hybrid functional B1-WC. [21] The electronic structure calculations were performed using the linear combination of atomic orbitals method as implemented in CRYSTAL first-principles code. [22] The used localized Gaussian-type basis sets which include polarization orbitals, are those from ref. [2] The $2 \times 2 \times 2$ $Fe_2TiSn$ full Heusler simple cubic supercells with 128 atoms/cell and Fm-3m symmetry (space group 225) were converted along 111 crystallographic direction to R3m symmetry (space group 160) after the explicit incorporation of the different considered defects. Brillouin zone integrations were performed using the following meshes of k-points: $4 \times 4 \times 4$ for simple cubic supercells, $12 \times 12 \times 12$ and $9 \times 9 \times 9$ for elemental bulk, binary and ternary compounds in Fe-V/Ti-Al/Sn(Sb) families used for the study of formation enthalpies. The self-consistent-field calculations were considered to be converged when the energy changes between interactions were smaller than $10^{-8}$ Hartree. An extralarge predefined pruned grid consisting of 75 radial points and 974 angular points was used for the numerical integration of charge density. Full optimizations of the lattice constants and atomic positions were performed with the optimization convergence of $5 \times 10^{-5}$ Hartree/Bohr in the root-mean square values of forces and $1.2 \times 10^{-3}$ Bohr in the root-mean square values of atomic displacements. For the defects with $2 \times 2 \times 2$ simple cubic supercells, the atomic positions were relaxed at the theoretical lattice constants. The level of accuracy in evaluating the Coulomb and exchange series is controlled by five tolerance parameters ($10^{-ITOL_j}$, j=1-5).[22] The ITOL values used in our calculations are 7, 7, 7, 9, and 30.

## 3. Results

### 3.1. Measured electronic transport properties

The SEM and EDS characterizations of as cast and annealed samples indicate that phase purity and microstructural homogeneity improve after annealing at 900°C. The samples are composed of a matrix with embedded traces of secondary phases. Good phase purity and microstructural homogeneity are observed in all the samples, except for the x = 0.6 sample, where the matrix is unmixed into two phases with different Sn:Sb compositions. The elemental content of the matrices are compatible with the nominal ones within few percent for all the samples $0 \leq x \leq 0.5$, confirming the correct stoichiometry as well as the correct substitution of Sb at Sn site; this is not the case for the x = 0.6 sample, where the Sn:Sb composition of the two matrix phases is nearly 20% off-stoichiometric. A table with detailed EDS results and SEM images of some samples are presented in Supporting Information, Table S1. Good phase purity is also confirmed by XRPD spectra, shown in Figure 1 (upper panel). The amounts of secondary phases are evaluated through Rietveld analysis, except for the x=0.6 sample, whose compositional inhomogeneity visible from EDS data made the Rietveld analysis unreliable. The hexagonal (P63/mmc) $Fe_2(Sn,Ti)$ secondary phase, a ternary phase derived from the binary $Fe_2Ti$ one, is present in all the samples, in very small percentages (1.4(1)%, 7.7(5)%, 5.9(6)%, and 5(1)% in the x=0, x=0.1, x=0.2, and x=0.5 samples respectively). Titanium is also present as a secondary phase in all the samples, in different percentages (0.60(6)%, 1.11(2)%, 0.54(8)%, and 0.42(6)% in the x=0, x=0.1, x=0.2, and

x=0.5 samples respectively) and with a cell parameter that slightly increases with increasing Sb, possibly due to Fe impurities. The cubic lattice parameter of the main phase (Fm-3m) is also extracted from the shift of Bragg reflections. The Vegard-like linear dependence of this lattice parameter with doping x, shown in the lower panel of Figure 1, confirms the expected substitution of Sb at Sn site.

Electrical conductivity $\sigma$ curves from 10 K to 380 K are presented in Figure 2a). The undoped sample is metallic, with an overall conductivity variation in the whole temperature range as small as 17%, typical of a semimetal, similarly to the literature data reported on samples annealed at lower temperatures. [3,19,20] The 10% doped sample exhibits metallic conductivity below 260 K and a transition to semiconducting behavior at higher temperature. In the 20%, 50% and 60% Sb doped samples, the temperature dependence is semiconducting in the whole temperature range, again in line with the literature data mentioned above. [19,20] In all cases, the change of conductivity in the temperature range 10 K - 380 K is very small, within 40%. A general trend of the absolute value of conductivity to decrease with increasing Sb doping is clearly observed, with $\sigma$ in the range $4 \times 10^2$ to $3 \times 10^3$ $\Omega^{-1} cm^{-1}$. However, absolute values of conductivity must be considered with care due to the ~15% experimental uncertainty on the geometrical factor.

The linear slope of the Hall resistance versus field in Hall effect measurements (Hall effect curves for selected representative samples are shown in Supporting Information, Figure S2) indicates that in all the samples, carriers are holes. The hole carrier density extracted in a single band approximation from the ordinary part of the Hall effect decreases slightly and monotonically with increasing Sb doping and it is ~$1.5$-$3 \times 10^{21}$ $cm^{-3}$ for doping up to 50% and ~$5 \times 10^{20}$ $cm^{-3}$ for 60% doping at room temperature, as shown in Figure 2b). In the undoped sample there is no anomalous Hall effect, while in the doped samples there is an anomalous Hall effect originating form magnetic ordering, well visible up to room temperature, whose magnitude increases with increasing doping, as also observed in ref. [19] Magnetoresistance is always negative except for the low doped samples $x \leq 0.2$ above room temperature (magnetoresistance curves for selected representative samples are shown in Supporting Information, Figure S3) and its magnitude increases with increasing doping up to $x = 0.5$, remaining however as small as few percent. Considering the presence of magnetic ordering evidenced by the anomalous Hall effect, it is plausible to assume that the negative magnetoresistance originates in scattering by spin fluctuations: namely, with increasing field, the magnetic moments get more and more aligned, and the fluctuations are increasingly suppressed, so that scattering by spin fluctuations decreases and the resistance decreases with increasing field. Figure 2c) shows Hall mobilities, all in the range of few $cm^2$ $V^{-1}$ $s^{-1}$. The largest mobility is observed in the most doped sample $x = 0.6$, possibly consistent with the finding of ref. [19] that substitution of Sn with Sb decreases further the antisite disorder. The increase of mobility with n-type Sb doping may originate from the presence of other $V_{Fe}$ and antisite defects with lower concentration besides $Sb_{Sn}$ scattering centers, and also from a better electronic screening of heavier holes. In the usual behavior, the mobility increases with the reduction of carrier concentration, and this is the overall trend of our samples. Within this general trend, some variations are possible due to the fact that different types of atomic defects, of which $V_{Fe}^{2-}$ are strong scattering centers for carriers, are simultaneously present in the samples.

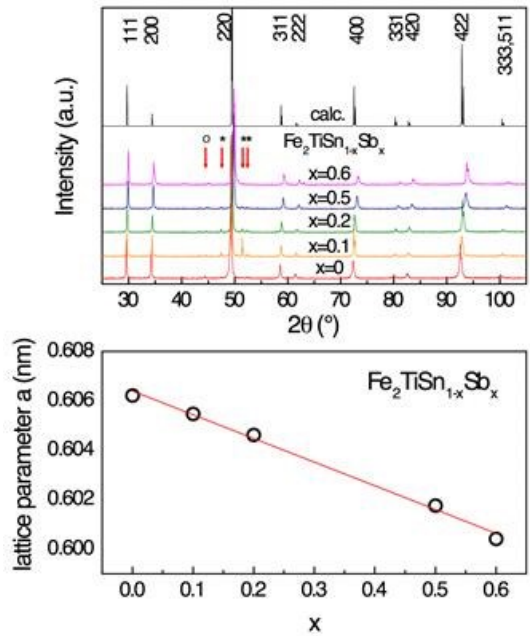

**Figure 1.** Upper panel: x-rays diffraction patterns of the $Fe_2TiSn_{1-x}Sb_x$ (x = 0.0, 0.1, 0.2, 0.5 and 0.6) samples. The calculated pattern and the Bragg reflections of the secondary phases are indicated ("*" for $Fe_2(Ti,Sn)$ and "$^O$" for Ti). Lower panel: cubic lattice parameter as a function of Sb doping x.

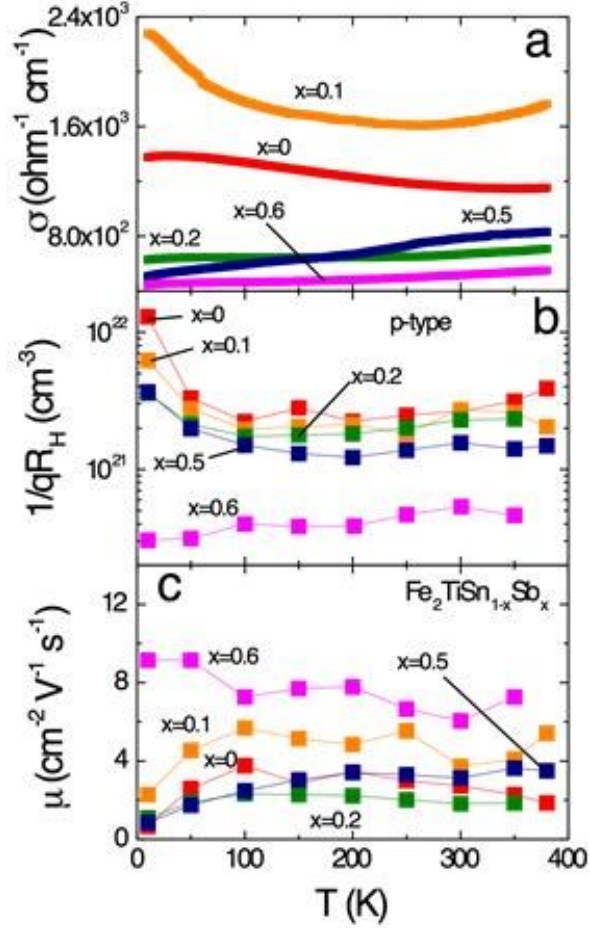

**Figure 2.** Electrical conductivity (a), carrier density (b) and carrier mobility (c) curves of $Fe_2TiSn_{1-x}Sb_x$ (x = 0.0, 0.1, 0.2, 0.5 and 0.6) samples.

As seen in Figure 3a), in all the samples, the Seebeck coefficient curves exhibit a nearly monotonic increase with increasing temperature, as expected from a non-gapped band structure and the sign of the Seebeck coefficient is positive, as expected from p-type carriers. Seebeck coefficient values are significantly larger than our published data on samples prepared by arc melting and annealed at 700°C [20] and even somewhat larger than data on samples annealed at 850°C. [3] Specifically around room temperature S values reach 56 μV/K in the 20% doped sample, 53 μV/K in the 10% and 50% doped samples and around 50 μV/K in the undoped and 60% doped samples. Most interesting, the thermoelectric power factors $S^2\sigma$ displayed in Figure 3b) reach high room temperature values, up to $4.8\times10^{-4}$ W $K^{-2}$ $m^{-1}$ in the 10% doped sample, four times larger than the largest value on Sb-doped samples annealed at 700°C. [20]

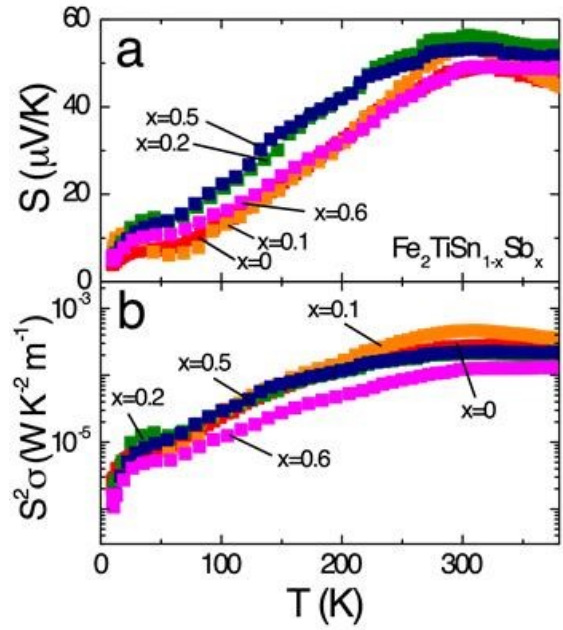

**Figure 3.** Seebeck coefficient (a) and power factor (b) curves of $Fe_2TiSn_{1-x}Sb_x$ (x = 0.0, 0.1, 0.2, 0.5 and 0.6) samples.

### 3.2. Calculated formation energies of defects and electronic properties

The formation energy ($\Delta H_{D,q}$) of a point defect D in charge state q is calculated as: [23]

$$\Delta H_{D,q} = ( E_{D,q} - E_H ) + \sum_i n_i \mu_i + qE_F + E_{corr} ,\qquad (1)$$

where $E_{D,q}$ is total energy of the host crystal containing a defect with charge state $q$, $E_H$ is total energy of the host crystal without defects, $n_i$ indicates the number of atoms of element $i$ that were added ($n_i < 0$) or removed ($n_i > 0$) form the defect, $\mu_i$ is chemical potential of element $i$, $E_F$ is Fermi energy referenced to the valence band maximum (*VBM*), and $E_{corr}$ are corrections including image charge correction ($E_l$) due to finite size effects of defect calculations performed using periodic supercells and electrostatic potential alignment correction ($E_{pa}$) for the charged supercells. $E_{pa} = q\Delta V_{pa}$, where $\Delta V_{pa}$ is potential alignment between the defect with charge $q$ and the host calculations. [23] The elemental chemical potential $\mu_i$ can be expressed relative to the reference elemental phase as $\mu_i = \mu_i{}^0 + \Delta\mu_i$, where $\mu_i{}^0$ is the chemical potential of element $i$ which forms in the reference ground state under standard conditions, and $\Delta\mu_i$ is the deviation from the reference chemical potential ($\Delta\mu_i = 0$ corresponds to element $i$ rich condition). The $\mu_i{}^0$ and $\Delta\mu_i$ values were estimated using FERE approach [24] (see Supporting Information for details).

The formation energies, $\Delta H_{D,q}$ of vacancies and Sb substitution at Fe, Ti and Sn sites of $Fe_2TiSn$, $V_{Fe}$, $V_{Ti}$, $V_{Sn}$, $Sb_{Sn}$, $Sb_{Fe}$, and $Sb_{Ti}$, estimated for different charged $q$ states and points from the phase stability region (Figure 4) at zero Fermi energy ($E_F = 0$) are shown in Table 1, whereas $\Delta H_{D,q}$ of single antisite defects in Table 2. For the most dominant vacancy and antisite defects, and $Sb_{Sn}$ ($Sb_{Sn}{}^{1-}$, $Sb_{Sn}{}^{1+}$) substitutions, we show in Figure 5 their Fermi energy dependence in (0, 1.5eV) energy range spanning from VBM to ~0.5eV above of conduction band minimum (CBM) of $Fe_2TiSn$. In the case of Fe-rich and Ti-rich conditions (A and B points from phase stability region (Figure 4)), the lowest formation energy has Ti at Sn, $Ti_{Sn,}$ antisite defect which is non-magnetic, and it increases the band gap by ~0.05 eV. We estimated a ~0.11eV band gap increase for two uncorrelated $Ti_{Sn}$ defects per supercell. This unexpected band gap increase is similar to that of Al/V inversion ($Al_V$ and $V_{Al}$ coupled antisite defects) in $Fe_2VAl$.[25] In the limit of full Ti substitution at Sn site, $Fe_2TiSn$ transforms to FeTi stable structure with the space group Pm-3m. As a result for Ti-rich and Fe-rich conditions $Ti_{Sn}$ antisite has stable formation energy ($\Delta H_{D,q} < 0$). For Fe-rich condition, $Fe_{Sn}$ ($Fe_{Sn}^{2-}$, $Fe_{Sn}^{1-}$, $Fe_{Sn}^{1+}$, $Fe_{Sn}^{2+}$) are magnetic defects with low or stable $\Delta H_{D,q}$. At Sn-, Sb-rich (E point) and Ti-, Sn-rich (F point) conditions, the lowest defect formation energies have $V_{Fe}$ ($V_{Fe}^{2-}$, $V_{Fe}^{2+}$), and $Ti_{Fe}^{1+}$ defects. $V_{Fe}^{2-}$ is the dominant nonmagnetic defect responsible for the formation of p-type intrinsic doping in $Fe_2TiSn$. These theoretical results are in agreement with the experimental evidence of $V_{Fe}$ in $Fe_2TiSb$ compound and the better suggested stoichiometric $Fe_{1.5}TiSb$ composition.[26] They are also consistent with our EDS analysis, which indicates Fe deficiency for all the samples except the x=0.1 one. The favourable Sb doping is at Sn site for all types of growing conditions. Also, this theoretical result is consistent with our EDS analysis, which confirms the expected Sb substitution, i.e. the presence of the expected amount of $Sb_{Sn}$ dopants. The $Sb_{Sn}$ ($Sb_{Sn}^{1-}$, $Sb_{Sn}^{1+}$) substitutions have very low or stable formation energies ($\Delta H_{D,q} < 0$), since $Fe_2TiSb$ and $Fe_{1.5}TiSb$ compounds are stable. $Sb_{Sn}^{1+}$ is a nonmagnetic defect acting as an n-type dopant. $Sb_{Sn}^{1-}$ is a magnetic defect acting as a p-type dopant, which becomes more dominant ($\Delta H_{D,q} < 0$) with increasing Sb doping and together with $V_{Fe}^{2-}$ may limit the decrease of p-type carrier concentration giving their nonmonotonic behavior in the Fe-poor growing conditions (F and E points).[20] On the other hand, in Fe-rich (A point) and Ti-rich (B point) growing conditions $Sb_{Sn}$ type defect have $\Delta H_{D,q} < 0$ and with increasing Sb doping at Sn site the n-type doping of $Fe_2TiSn$ could be achieved also through $Sb_{Sn}$ contribution. For very high Sb doping concentrations we expect a pseudogap formation in the spin up channel, which is suggested from the electronic density of states (DOS) analysis of $Sb_{Sn}$ and $Sb_{Sn}^{1-}$ (Figure 6g and h).

DOS of $V_{Fe}$ ($V_{Fe}^{2-}$), the most dominant antisite defects $Ti_{Fe}^{1+}$, $Ti_{Sn}$, $Fe_{Sn}^{2-}$, $Fe_{Sn}^{1+}$, and $Sb_{Sn}$ ($Sb_{Sn}^{1-}$, $Sb_{Sn}^{1+}$) is shown in Figures 6 and S5. $V_{Fe}$ introduces localized in-gap electronic states, and it has a magnetic moment of 2 $m_B$ per defect (Figure 6a). $V_{Fe}^{2-}$ does not significantly change the electronic states near $Fe_2TiSn$ band gap giving a nonmagnetic semiconducting behavior with a band gap of ~1.0 eV (Figure 6b). $Ti_{Fe}^{1+}$ and $Fe_{Sn}^{1+}$, are magnetic defects acting as n-type dopants having the magnetic moments of ~1.7 $\mu_B$ and ~3.7 $\mu_B$ per defect, respectively (Figure 6c, f). $Fe_{Sn}^{2-}$ is a magnetic defect acting as p-type dopant with a magnetic moment of ~6.7 $\mu_B$ per defect (Figure 6e). All these magnetic antisite defects introduce deep in-gap states, which affect the electronic properties near the band gap of $Fe_2TiSn$. When such defects are present at concentration comparable with the experimental values, they are significantly affecting the electronic properties giving a metallic behavior with large DOS at Fermi energy, or semimetallic behavior with the formation of a pseudogap. Optical spectroscopy experiments show the presence of a pseudogap of ~0.87 eV in $Fe_2TiSn$.[27] $Ti_{Sn}$, is a nonmagnetic defect which increases the theoretical band gap of $Fe_2TiSn$ from 1.04eV to 1.09eV per defect (Figure 6d). $Sb_{Sn}$ shows half metallic ferromagnetic behavior with a magnetic moment of 1 $\mu_B$. The metallic in-gap states for spin up component are spread to ~0.5 eV energy range close to middle of $Fe_2TiSn$ band gap. $Sb_{Sn}^{1-}$ type defect has a magnetic moment of 2 $\mu_B$ and introduces resonant states near VBM and CBM of $Fe_2TiSn$ decreasing the band gap to ~0.52 eV (resp. ~0.93 eV) for spin up (resp. spin down) channel. $Sb_{Sn}^{1+}$ type defect is nonmagnetic and does not significantly change the electronic states near $Fe_2TiSn$ band gap.

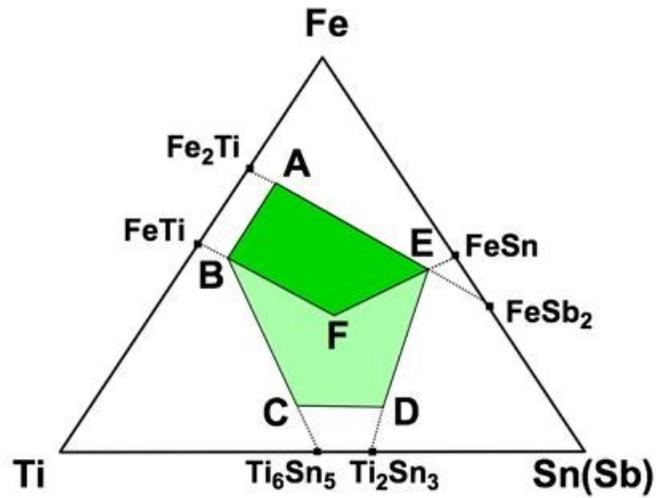

**Figure 4.** Phase stability of $Fe_2TiSn(Sb)$ full Heusler compounds. The region with favorable stable conditions is marked by ABFE area in light green color, which excludes the formation of ternary FeTiSb and FeTiSn hexagonal phases.

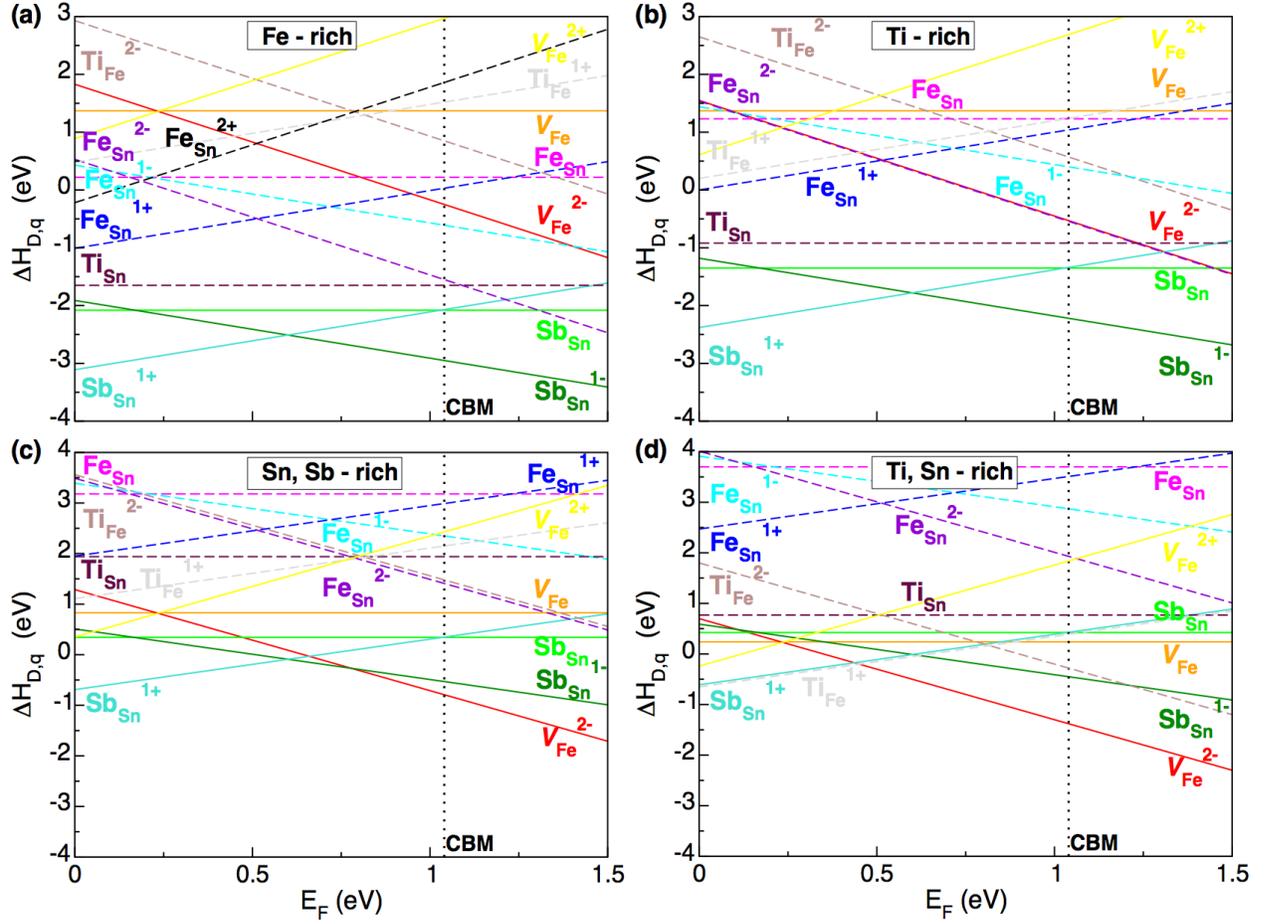

**Figure 5.** Fermi energy $E_F$ dependence of formation energy $\Delta H_{D,q}$ for $V_{Fe}$ ($V_{Fe}^{2-}$, $V_{Fe}^{2+}$) vacancies, $Fe_{Sn}$ ($Fe_{Sn}^{2-}$, $Fe_{Sn}^{1-}$, $Fe_{Sn}^{1+}$), $Ti_{Fe}^{2-}$, $Ti_{Fe}^{1+}$, and $Ti_{Sn}$ single antisite defects, and $Sb_{Sn}$ ($Sb_{Sn}^{1-}$, $Sb_{Sn}^{1+}$) substitutions at (a) Fe-rich (A point from Figure 4), (b) Ti-rich (B point from Figure 4), (c) Sn-, and Sb-rich (E point from Figure 4), and (d) Ti-, and Sn-rich (F point from Figure 4) conditions. $E_F = 0$ represents the valence band maximum VBM and the conduction band minimum CBM is shown in dashed line for $Fe_2TiSn$.

**Table 1.** Formation energy, $\Delta H_{D,q}$ of vacancies $V$ and Sb substitutions at Sn, Fe and Ti sites, $V_{Sn}$, $V_{Fe}$, $V_{Ti}$, $Sb_{Sn}$, $Sb_{Fe}$, and $Sb_{Ti}$ estimated at different points from the phase stability region (Fig-

ure 4) and at zero Fermi energy ($E_f$=0) corresponding to VBM of $Fe_2TiSn$. $E_{corr}$ for charged $q$ states and total magnetic moment per supercell $M_{tot}$ are also included.

| | $\Delta H_{D,q}$ (eV/atom) | | | | | | $E_{corr}$ (eV/atom) | $M_{tot}$ ($\mu_B$) |
|---|---|---|---|---|---|---|---|---|
| | A | B | C | D | E | F | | |
| $V_{Fe}$ | 1.37 | 1.09 | 0.73 | 0.72 | 0.83 | 0.24 | | 2 |
| $V_{Fe}^{2-}$ | 1.83 | 1.55 | 1.19 | 1.18 | 1.29 | 0.70 | + 0.46 | 0 |
| $V_{Fe}^{1-}$ | 1.62 | 1.34 | 0.98 | 0.97 | 1.08 | 0.49 | + 0.25 | 1 |
| $V_{Fe}^{2+}$ | 0.89 | 0.61 | 0.25 | 0.24 | 0.35 | -0.24 | - 0.48 | 1.92 |
| $V_{Ti}$ | 7.04 | 7.04 | 7.04 | 6.08 | 5.87 | 7.04 | | 0 |
| $V_{Ti}^{4-}$ | 7.48 | 7.48 | 7.48 | 6.52 | 6.31 | 7.48 | + 0.44 | 0 |
| $V_{Ti}^{3-}$ | 7.65 | 7.65 | 7.65 | 6.69 | 6.48 | 7.65 | + 0.61 | 1 |
| $V_{Ti}^{2-}$ | 7.34 | 7.34 | 7.34 | 6.38 | 6.17 | 7.34 | + 0.30 | 1 |
| $V_{Ti}^{1-}$ | 6.94 | 6.94 | 6.94 | 5.98 | 5.77 | 6.94 | - 0.1 | 0 |
| $V_{Sn}$ | 3.81 | 4.54 | 5.26 | 6.23 | 6.23 | 6.23 | | 4 |
| $V_{Sn}^{4-}$ | 4.33 | 5.06 | 5.78 | 6.75 | 6.75 | 6.75 | + 0.52 | 0 |
| $V_{Sn}^{3-}$ | 4.02 | 4.75 | 5.47 | 6.44 | 6.44 | 6.44 | + 0.21 | 1 |
| $V_{Sn}^{2-}$ | 4.07 | 4.80 | 5.52 | 6.49 | 6.49 | 6.49 | + 0.26 | 1.94 |
| $V_{Sn}^{1-}$ | 4.00 | 4.73 | 5.45 | 6.42 | 6.42 | 6.42 | + 0.19 | 3 |
| $Sb_{Sn}$ | -2.08 | -1.35 | -0.63 | 0.34 | 0.34 | 0.42 | | 1 |
| $Sb_{Sn}^{1-}$ | -1.91 | -1.18 | -0.46 | 0.51 | 0.51 | 0.59 | + 0.17 | 2 |
| $Sb_{Sn}^{1+}$ | -3.11 | -2.38 | -1.66 | -0.69 | -0.69 | -0.61 | - 1.03 | 0 |
| $Sb_{Fe}$ | 4.74 | 4.46 | 4.10 | 4.09 | 4.20 | 3.69 | | 3 |
| $Sb_{Fe}^{3-}$ | 5.38 | 5.10 | 4.74 | 4.73 | 4.84 | 4.33 | + 0.64 | 6 |
| $Sb_{Fe}^{1-}$ | 5.14 | 4.86 | 4.50 | 4.49 | 4.60 | 4.09 | + 0.40 | 2 |
| $Sb_{Fe}^{1+}$ | 4.45 | 4.17 | 3.81 | 3.80 | 3.91 | 3.40 | - 0.29 | 0 |
| $Sb_{Ti}$ | 4.03 | 4.03 | 4.03 | 3.07 | 2.86 | 4.11 | | 0 |

**Table 2.** Formation energy, $\Delta H_{D,q}$ of single antisite defects estimated at different points from the phase stability region (Figure 4) and at zero Fermi energy ($E_F$=0) corresponding to VBM of $Fe_2TiSn$. $E_{corr}$ for charged $q$ states and total magnetic moment per supercell $M_{tot}$ are also included.

| | $\Delta H_{D,q}$ (eV/atom) | | | | | | $E_{corr}$ (eV/atom) | $M_{tot}$ ($\mu_B$) |
|---|---|---|---|---|---|---|---|---|
| | **A** | **B** | **C** | **D** | **E** | **F** | | |
| $Fe_{Ti}$ | 4.09 | 4.37 | 4.73 | 3.78 | 3.46 | 5.22 | | 0 |
| $Ti_{Fe}$ | 1.61 | 1.33 | 0.97 | 1.92 | 2.24 | 0.48 | | 2 |
| $Ti_{Fe}^{1-}$ | 1.83 | 1.55 | 1.19 | 2.14 | 2.46 | 0.70 | + 0.22 | 3 |
| $Ti_{Fe}^{2-}$ | 2.93 | 2.65 | 2.29 | 3.24 | 3.56 | 1.80 | + 1.32 | 0 |
| $Ti_{Fe}^{1+}$ | 0.48 | 0.20 | -0.16 | 0.79 | 1.11 | -0.65 | - 1.13 | 1.71 |
| $Ti_{Fe}^{2+}$ | 1.48 | 1.20 | 0.84 | 1.79 | 2.11 | 0.35 | - 0.13 | 0 |
| $Fe_{Sn}$ | 0.22 | 1.23 | 2.31 | 3.29 | 3.18 | 3.7 | | 4.72 |
| $Fe_{Sn}^{1-}$ | 0.43 | 1.44 | 2.52 | 3.50 | 3.39 | 3.91 | + 0.21 | 1.77 |
| $Fe_{Sn}^{2-}$ | 0.53 | 1.54 | 2.62 | 3.60 | 3.49 | 4.01 | + 0.31 | 6.74 |
| $Fe_{Sn}^{1+}$ | -1.01 | 0.0 | 1.08 | 2.06 | 1.95 | 2.47 | - 1.23 | 3.74 |
| $Fe_{Sn}^{2+}$ | -0.22 | 0.79 | 1.87 | 2.85 | 2.74 | 3.26 | - 0.44 | 2.74 |
| $Sn_{Fe}$ | 7.24 | 6.23 | 5.15 | 4.17 | 4.28 | 3.69 | | 2 |
| $Sn_{Fe}^{2-}$ | 8.05 | 7.04 | 5.96 | 4.98 | 5.09 | 4.50 | +0.81 | 2 |
| $Sn_{Fe}^{2+}$ | 7.10 | 6.09 | 5.01 | 4.03 | 4.14 | 3.55 | - 0.14 | 0 |
| $Sn_{Ti}$ | 5.61 | 4.88 | 4.16 | 2.23 | 2.02 | 3.19 | | 0 |
| $Ti_{Sn}$ | -1.65 | -0.92 | -0.20 | 1.73 | 1.94 | 0.77 | | 0 |

## 4. Discussion

We first address the comparison of our results with those of our previous work, carried out on $Fe_2TiSn$ samples prepared by arc melting, annealed at different temperatures up to 800°C and doped with Sb up to 20% on the Sn site. [20] In all the samples of this work and of our previous work, [20] p-type transport and semimetallic or weakly semiconducting behavior is observed. As demonstrated by the previous [20] and present theoretical calculations, this is the result of antisite and vacancy defects, which totally overshadow the intrinsic behavior that was instead predicted to yield n-type power factor values over $10x10^{-3}$ W m$^{-1}$ K$^{-2}$ in Sb doped samples. [2]

Regarding the preparation protocol, in our previous work, [20] a trend of increasing mobility, Seebeck coefficient and power factor was observed with increasing annealing temperature $T_{AN}$ up to 800°C, however also a larger amount of secondary phases was detected with increasing $T_{AN}$. In this work, with the high-frequency melting preparation method, a good matrix stoichiometry is detected even with $T_{AN}$ as high as 900°C, and at the same time this higher annealing temperature favours grain formation and improves phase homogeneity. As a consequence, the samples of this work exhibit larger S in the range 49-56 µV/K at room temperature and larger power factors up to $4.8x10^{-4}$ W K$^{-2}$ m$^{-1}$ at room temperature, closer to intrinsic behavior. The values of S

~50 μV/K are typical for semimetals with a pseudogap and small DOS at the Fermi energy. This is clear evidence that the samples of previous work have a more metallic behavior (larger DOS at Fermi energy) and a larger antisite disorder than the samples of the present work. The stronger antisite disorder ($Ti_{Fe}^{1+}$, $Fe_{Sn}^{2-}$, $Fe_{Sn}^{1-}$, $Fe_{Sn}^{1+}$, and $Fe_{Sn}^{2+}$) in the undoped $Fe_2TiSn$ samples of previous work is consistent with the observation of negative magnetoresistance in those samples [20], that can be plausibly attributed to magnetic scattering.

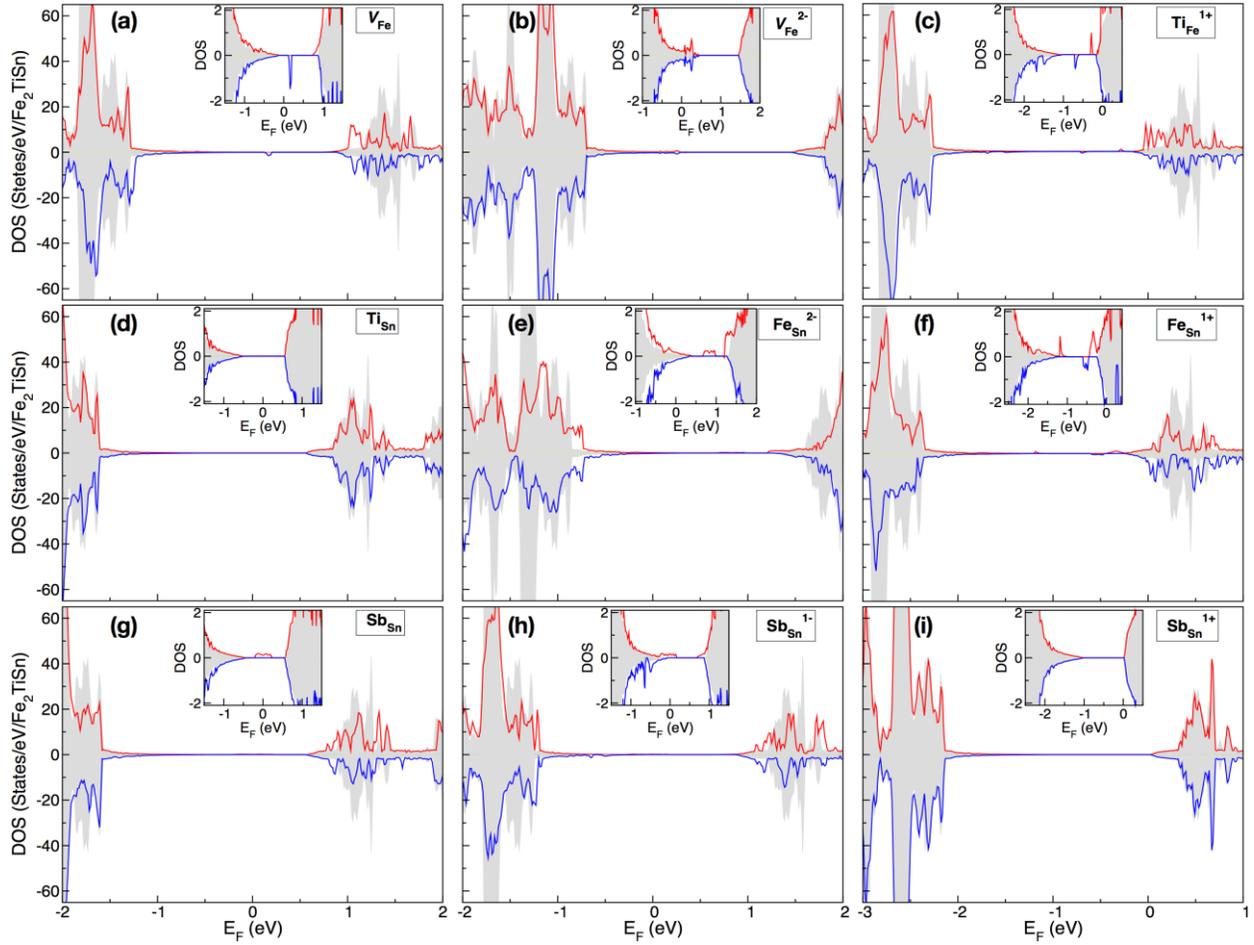

**Figure 6.** Electronic density of states (DOS) for spin up (red color) and spin down (blue color) channels of (a) $V_{Fe}$, (b) $V_{Fe}^{2-}$, (c) $Ti_{Fe}^{1+}$, (d) $Ti_{Sn}$, (e) $Fe_{Sn}^{2-}$, (f) $Fe_{Sn}^{1+}$, (g) $Sb_{Sn}$, (h) $Sb_{Sn}^{1-}$, and (i) $Sb_{Sn}^{1+}$. DOS of $Fe_2TiSn$ is shown in background gray colour. The electronic states near $Fe_2TiSn$ band gap are shown in the insets. $E_F=0$ represents the energy position of the different defects.

Regarding the effect of Sb doping, in the $Fe_2TiSn$ samples annealed at 700°C for 8 days of our previous work, [20] disorder related to Sb doping played a major role over charge filling. The slight improvement of the electric and thermoelectric properties in the 10% Sb doped sample was explained by features in the DOS induced by antisite and vacancy defects. As for the 20% Sb doped sample, it is possible that Anderson localization by disorder and charge compensation determined the eventual decrease in net carrier concentration, mobility, and thermoelectric power factor. On the contrary, in the present work, annealing at 900°C mitigates disorder and the effect of Sb doping is n-type charge filling and thus charge compensation. Indeed, while increasing Sb doping, carrier concentration decreases monotonically, and conductivity decreases and turns to a semiconducting temperature dependence. As for thermoelectric properties, the Seebeck coefficient first increases with Sb doping and then decreases with further doping, being maximum in the x=0.2 sample, 56 μV/K at room temperature. This non-monotonic behavior may result from the competition between different types of defects $V_{Fe}$ (p-type $V_{Fe}^{2-}$ dopant), and $Sb_{Sn}$ (p-type $Sb_{Sn}^{1-}$, and n-type $Sb_{Sn}^{1+}$ dopants), as indicated by our calculations (Figure 5). The power factor is also non monotonic with doping being maximum in the x = 0.1 sample, with a $4.8x10^{-4}$ W $m^{-1}$ $K^{-2}$ value around room temperature.

Our calculation, supported by EDS analysis, unveils the type of defects that determine p-type and n-type doping, as well as magnetism. The p-type doping is mainly associated with $V_{Fe}^{2-}$ and $Sb_{Sn}^{1-}$ charged defects in Fe-poor growing conditions, since antisite disorder is weaker in the samples of this work. On the other hand, the compensating n-type doping observed with increasing Sb doping is associated mainly with $Sb_{Sn}$ and charged $Sb_{Sn}^{1+}$ substitutional dopants, but also $Fe_{Sn}^{2+}$, $Fe_{Sn}^{1+}$ (Fe-rich growing condition), and $Ti_{Fe}^{1+}$ (Ti-, Sn-rich growing condition) antisites can contribute to n-type doping. The formation of $Sb_{Sn}^{1-}$ and $Sb_{Sn}^{1+}$ charged dopants is favourable in all synthesis conditions. The $Sb_{Sn}^{1-}$ and $Sb_{Sn}$ substitutions are mainly responsible for the magnetic behavior observed in the doped samples, evident from the anomalous Hall effect and negative magnetoresistance, but other $Ti_{Fe}^{1+}$, and charged $Fe_{Sn}$ magnetic defects can contribute depending on the growing conditions.

Finally, it is worth comparing our experimental and theoretical results with experimental data of literature. In Novitskii et al.,[28] $Fe_2TiSn$ samples prepared by mechanochemical synthesis of metal precursors followed by spark plasma sintering exhibit more metallic behavior and lower Seebeck coefficients than our samples and the authors attribute this transport behavior to the presence of Fe-Ti disorder ($Fe_{Ti}$ and $Ti_{Fe}$ antisite defects). The amount of secondary metallic phases in the samples of Novitskii et al.[28] is comparable with that of our doped samples (~5-8%), and much smaller than the value of ~1.4% for our undoped sample. Since all our samples show semimetallic behavior with the indication of a pseudogap, we do not expect the transport to be significantly affected by the secondary phases; instead, we rather suggest that the difference between our results and those of the Novitskii et al.[28] is coming from the fact that our samples have less antisite magnetic defects with metallic behavior, giving the improved thermoelectric performance. Other literature works presenting electric and thermoelectric properties of $Fe_2TiSn$ pre-pared by different methods (polycrystalline $Fe_{2-x}Ti_{1+x}Sn$ with Fe-poor and Fe-rich stoichiometry, prepared by solid state reaction and arc melting, [15] polycrystalline $Fe_2TiSn_{1-x}Si_x$ ($0 \leq x \leq 1$) prepared by induction melting, [29] by self-propagating high-temperature synthesis (SHS) followed by spark plasma sintering (SPS) [30]) report all p-type behavior and, in general, lower Seebeck coefficients. [15,29,30] According to our calculations, these data are explained with the presence of higher density of charged iron vacancies $V_{Fe}^{2-}$ and other antisite defects such as $Fe_{Sn}^{2-}$, in comparison with our samples. On the other hand, magnetism is observed in polycrystalline samples prepared by arc melting, [16,17,31,32] which can be explained by the presence of antisite defects such as $Fe_{Sn}$ and $Ti_{Fe}$.

## 5. Conclusions

In this work, we present electrical and thermoelectrical properties of Sb doped full-Heusler compounds, fabricated by high-frequency melting and optimized annealing conditions, namely annealing at 900°C for 13 days. Thanks to these controlled preparation conditions, we obtain samples with fairly good phase purity, homogeneous microstructure and good matrix stoichiometry. Mainly $V_{Fe}^{2-}$ vacancy defects and $Sb_{Sn}^{1-}$ charged dopants determine p-type transport in all the samples. However, with increasing n-type Sb doping, p-type charge carriers are progressively compensated, determining a lower net carrier density and a smaller conductivity indicating that the Fermi energy is displaced toward the pseudogap, as also calculated for the $Fe_2VAl$ system. [7] Thermoelectric properties reach their best values close to room temperature and are improved by Sb doping. The highest power factor $4.8 \times 10^{-4}$ W $m^{-1}$ $K^{-2}$ is obtained in the x = 0.1 sample and the highest Seebeck coefficient ~56 μV/K in the x = 0.2 sample. However, these p-type power-factor values are still by far smaller than theoretical n-type power factors predicted in $Fe_2TiSn_{1-x}Sb_x$ samples free of antisite and vacancy defects. In all types of growing conditions except Fe-rich condition, our calculations indicate that $V_{Fe}^{2-}$ and $Sb_{Sn}^{1-}$ charged defects are dominant being responsible for the p-type doping. Ti-rich is the most favourable growing condition, which reduces the formation of $V_{Fe}^{2-}$ defects and of charged magnetic detects ($Fe_{Sn}$ and $Ti_{Fe}$) and favours the formation of $Ti_{Sn}$ defect possessing a band gap opening effect. [5,6,7] This effect helps to preserve the semiconducting character of the electronic properties and to improve TE properties of $Fe_2TiSn$ by doping. In Ti-rich conditions and large Sb doping, our calculations predict a pseudogap formation in $Fe_2TiSn_{1-x}Sb_x$, like that of metastable thin-films of $Fe_2V_{0.8}W_{0.2}Al$ full Heusler alloys for which the very high PF's $\sim 20$–$40x10^{-3}$ W m$^{-1}$ K$^{-2}$ were recently measured. [33] This work shows that, even if ever present, atomic disorder in $Fe_2TiSn$ full-Heusler compounds plays a major role in the electrical and thermoelectrical transport properties and that a proper engineering of stoichiometry may yield enhanced thermoelectric properties.

**Corresponding Author**

Corresponding Author e-mail: Daniel.Bilc@uliege.be

**Author Contributions**

The manuscript was written through contributions of all authors. All authors have given approval to the final version of the manuscript.

‡ These authors (I.P. and D.I.B.) contributed equally.

**Funding Sources**

D.I.B. acknowledges support for the computational resources from the high performance computational facility of Babes-Bolyai University (MADECIP, POSCCE COD SMIS 48801/1862) co-financed by the European Regional Development Fund. Additional computational resources have been provided by the Consortium des Équipements de Calcul Intensif (CÉCI), funded by the Fonds de la Recherche Scientifique de Belgique (F.R.S.-FNRS) under Grant No. 2.5020.11 and by the Walloon Region.

**Supporting Information**

Electronic supporting information: Scanning electron microscopy and energy dispersive spectroscopy analyses (Table S1, Figure S1); Rietveld analysis (Figure S2); Hall effect and magnetoresistance curves (Figure S3-S4); Technical details for the estimation of chemical potentials; DOS analysis in a larger energy range (Figure S5).

# Supporting Information

# Roles of Defects and Sb-doping in the Thermoelectric Properties of Full-Heusler Fe$_2$TiSn

*Ilaria Pallecchi [a,‡], Daniel I. Bilc [*,b ‡], Marcella Pani [c,a], Fabio Ricci [d], Sébastien Lemal [d], Philippe Ghosez [d], Daniele Marré [e,a]*

[a] CNR-SPIN, Dipartimento di Fisica, Via Dodecaneso 33, 16146, Genova, Italy

[b] Faculty of Physics, Babeş-Bolyai University, 1 Kogalniceanu, RO-400084 Cluj-Napoca, Românîa

[c] Dipartimento di Chimica e Chimica Industriale, Università di Genova, Via Dodecaneso 31, I-16146 Genova, Italy

[d] Theoretical Materials Physics, Q-MAT, CESAM, Université de Liège (B5), B-4000 Liège, Belgium

[e] Dipartimento di Fisica, Università di Genova, Via Dodecaneso 31, I-16146 Genova, Italy

[*] Corresponding Author e-mail: Daniel.Bilc@uliege.be

# 1. Scanning electron microscopy and energy dispersive spectroscopy analyses

Table S1 reports energy dispersive spectroscopy (EDS) chemical characterization of as cast and annealed $Fe_2TiSn_{1-x}Sb_x$ samples and Figure S1 shows scanning electron microscopy (SEM) images of two representative compositions of as cast and annealed samples.

**Table S1:** EDS chemical characterization of as cast and annealed $Fe_2TiSn_{1-x}Sb_x$ samples

| $Fe_2TiSn_{1-x}Sb_x$ sample | Fe:Ti:Sn:Sb composition | |
|---|---|---|
| | As cast | Annealed at 900°C-13 days |
| x=0 | | **Global: 49.5 : 25.1 : 25.4 : 0**<br>Matrix: 49.7 : 24.9 : 25.4 : 0<br>Spurious phases:<br>- Ti |
| x=0.1 | **Global: 49.1 : 25.5 : 23.0 : 2.5**<br>Matrix: 49.9 : 25.3 : 23.2: 1.6<br>Spurious phases:<br>- Fe₂(Ti,Sn) 64.9 : 26.1 : 9.0;<br>- FeTi(Sn,Sb) 33.7 : 31.0 : 21-24 : 11-15 (variable Sn:Sb);<br>- TiO₂ | **Global: 50.5 : 24 : 22.8 : 2.7**<br>Matrix: 50.6 : 23.4 : 23.7: 2.3<br>Spurious phases:<br>- TiO₂ |
| x=0.2 | **Global: 49.0 : 26.1 : 20.1 : 4.8**<br>Matrix: 50.2 : 25.5 : 20.8: 3.5<br>Spurious phases:<br>- Fe₂(Ti,Sn) 65.6 : 26.0 : 8.4<br>- FeTi(Sn,Sb) 33.7 : 30.6: 20.5 : 15.2<br>- Sn | **Global: 48.1 : 25.7 : 20.8 : 5.4**<br>Matrix: 49.0 : 25.4 : 21.2: 4.4<br>Spurious phases:<br>- Fe₂(Ti,Sn) 65.7 : 26.3 : 8 |
| x=0.5 | **Global: 48.8 : 25.5 : 12.9 : 12.8**<br>Matrix: 52.8 : 22.2 : 15.2: 9.7<br>Spurious phases:<br>- Fe₂(Ti,Sn,Sb) 65.4 : 25.4 : 7.6 : 1.6<br>- FeTi(Sn,Sb): 33.1 : 30.1 : 11.4 : 25.4; this phase forms an eutectic with Fe₂(Sn,Sb)<br>- Sn at grain boundaries | **Global: 48.6 : 25.8 : 12.8 : 12.8**<br>Matrix: 48.4 : 25.5 : 13.5: 13<br>Spurious phases:<br>- FeTi(Sn,Sb): 44.4 : 27.3 : 8.1 : 20.2<br>- Fe₂(Ti,Sn,Sb) 65.5 : 25.5 : 7.7 : 1.3, even as eutectic |
| x=0.6 | **Global: 49.6 : 24.5 : 10.1 : 15.8**<br>Matrix: 53.3 : 22. : 12.5: 12.2<br>- Fe₂(Ti,Sn,Sb) 65.8 : 24.4 : 7.6 : 2.1; this phase forms an eutectic with the matrix<br>- FeTi(Sn,Sb) 34.4 : 28.7 : 7.7 : 29.2 | **Global: 48.1 : 24.5 : 11.1 : 16.3**<br>Matrix1: 49.4 : 23.8: 12.1 : 14.7<br>Matrix2: 45.4 : 26.2 : 7.7: 20.8 |

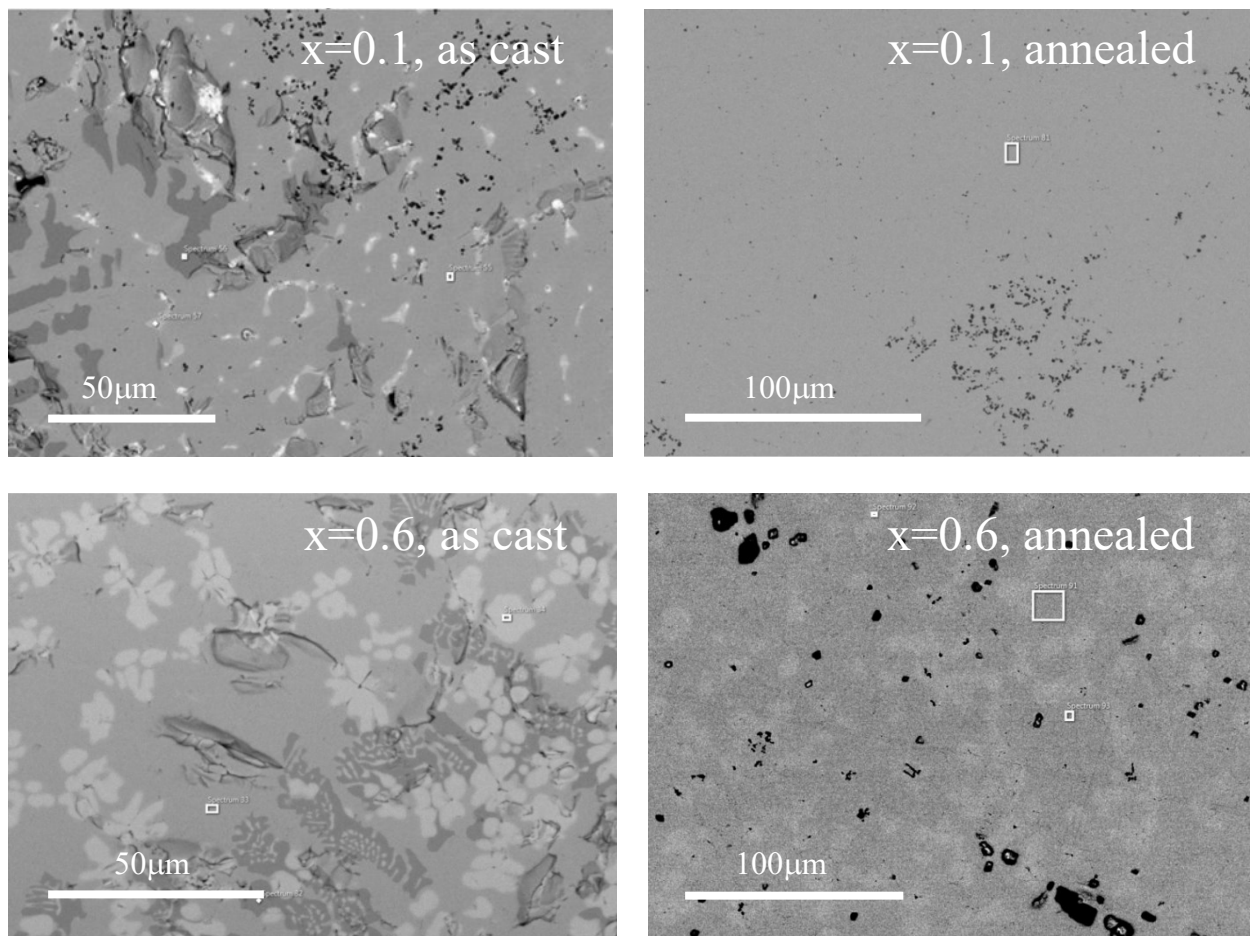

**Figure S1:** SEM images of some as cast and annealed Fe₂TiSn₁₋ₓSbₓ samples.

## 2. Rietveld analysis

Rietveld analysis was carried out to estimate the amount of secondary phases for all the samples, except for the x=0.6 sample, whose compositional inhomogeneity visible from EDS data made the Rietveld analysis unreliable. The hexagonal (P6₃/mmc) Fe₂(Sn,Ti) secondary phase, a ternary phase derived from the binary Fe₂Ti one, is present in all the samples, in small percentages (1.4(1)%, 7.7(5)%, 5.9(6)%, and 5(1)% in the x=0, x=0.1, x=0.2, and x=0.5 samples, respectively). Titanium is also present as a secondary phase in all the samples, in very small percentages (0.60(6)%, 1.11(2)%, 0.54(8)%, and 0.42(6)% in the x=0, x=0.1, x=0.2, and x=0.5 samples, respectively) and with a cell parameter that slightly increases with increasing Sb, possibly due to incorporation of Fe in the Ti lattice. In Figure S2, the Rietveld plots are shown.

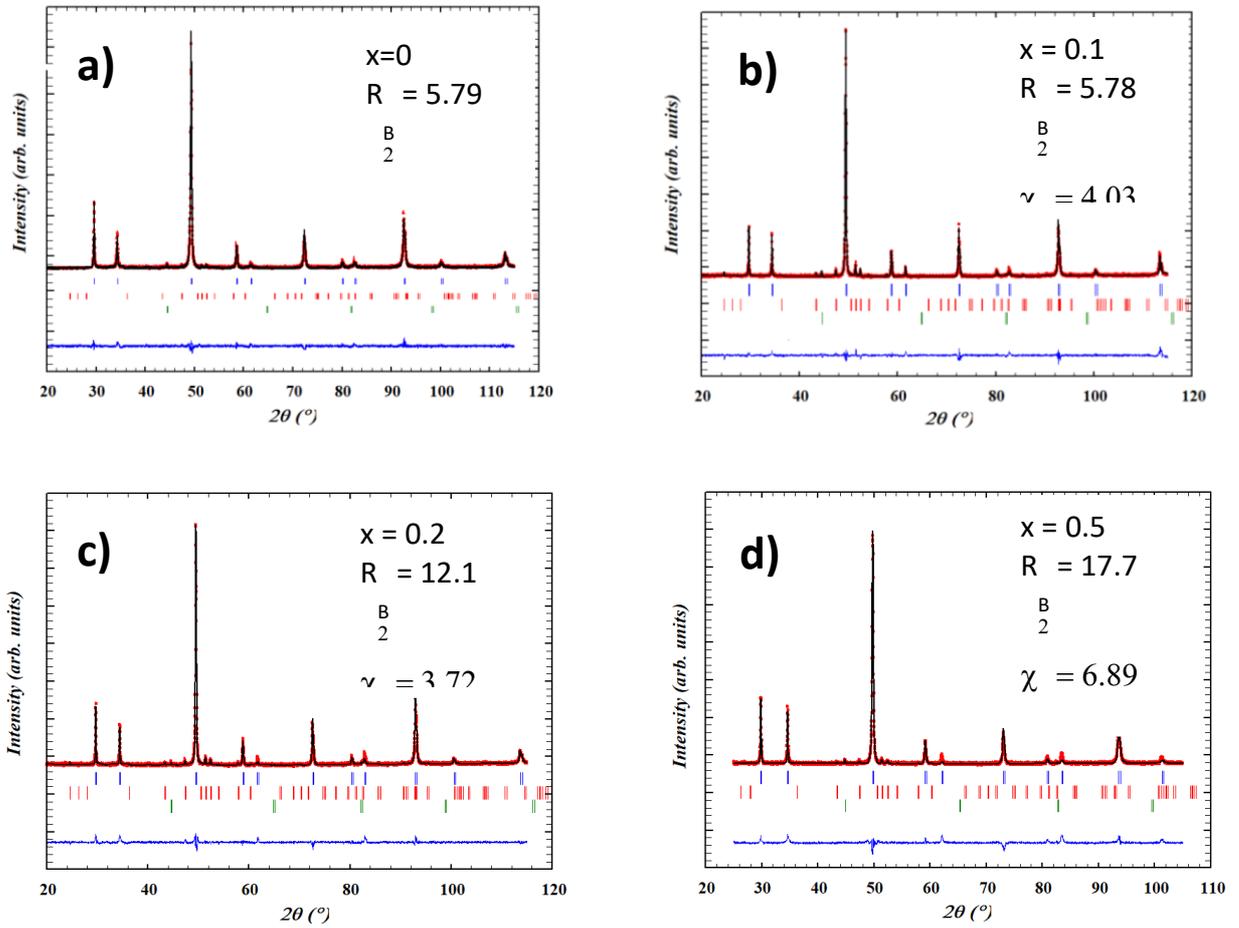

**Figure S2:** Rietveld plots of samples x=0 (a), x=0.1 (b), x=0.2 (c), x=0.5 (d). Experimental points, calculated profiles and difference ($Y_{obs}$-$Y_{calc}$) curves are represented in red, black and blue, respectively. Vertical bars correspond to the positions of calculated Bragg peaks; from top to bottom, they represent the Heusler phase, the $Fe_2(Ti,Sn)$ phase and Ti. The $R_B$ values reported in each plot are referred to the main Heusler phase.

### 3. Hall effect and magnetoresistance curves

Hall effect measurements indicate that in the $Fe_2TiSn_{1-x}Sb_x$ (x=0) sample only an ordinary Hall contribution, linearly dependent on the applied field, is present. On the other hand, in Sb doped ($0.1 \leq x \leq 0.6$) samples, an anomalous Hall contribution originating form magnetic ordering is well visible up to room temperature, and its magnitude increases with increasing doping. From the linear slope of the ordinary contribution, it is extracted that carriers are holes and their concentrations $n$ monotonically decrease with increasing Sb doping, as a consequence of charge compensation. The concentration $n$ ranges from few times $10^{21}$ cm$^{-3}$ for doping up to 50% and few times $10^{20}$ cm$^{-3}$ for 60% doping, as shown in Figure 1b in the main text) of the main text. In Figure S3, representative Hall effect curves of x=0.2 and x=0.5 samples are presented.

Magnetoresistance, defined as the variation of longitudinal resistance $R_{xx}$ with field, normalized to the zero-field value $(R_{xx}(H)-R_{xx}(H=0))/R_{xx}(H=0)$, is negative, small in magnitude (up to few percent at H= 90000 Oe) and decreasing with increasing temperature. Its magnitude increases monotonically with doping up to 0.05 in the x=0.5 samples and decreases again to 0.03 in the x=0.6 sample. Only for doping $x \leq 0.2$ and at the highest temperatures $T \geq 350K$, magnetoresistance is positive and proportional to the squared field $\propto H^2$, according to the semiclassical cyclotron picture. The negative magnetoresistance term has likely a magnetic origin, consistent with the above described presence of anomalous Hall effect. In Figure S4, representative magnetoresistance curves of x=0.2 and x=0.5 samples are shown.

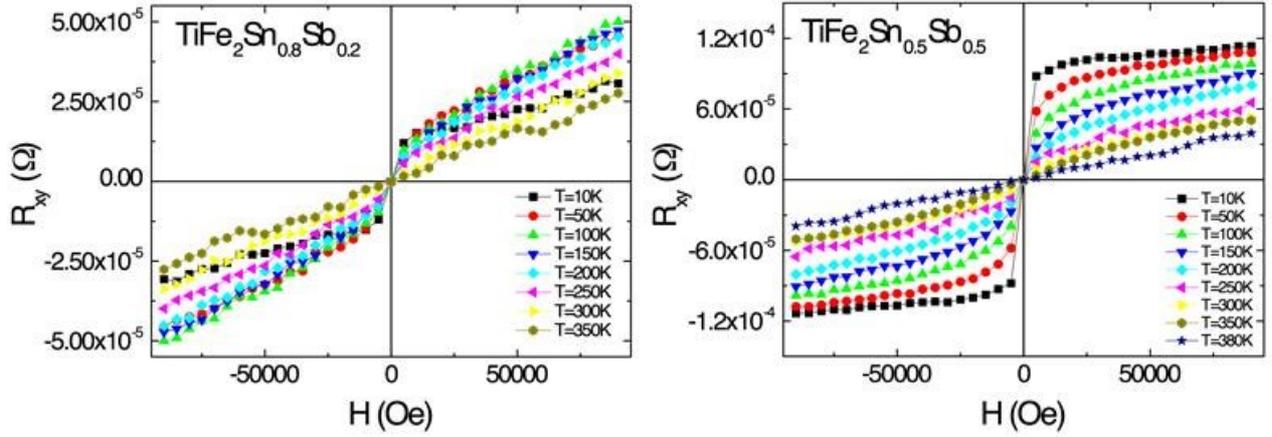

**Figure S3:** Transverse resistance $R_{xy}$ curves versus applied field, measured in the x=0.2 and x=0.5 samples.

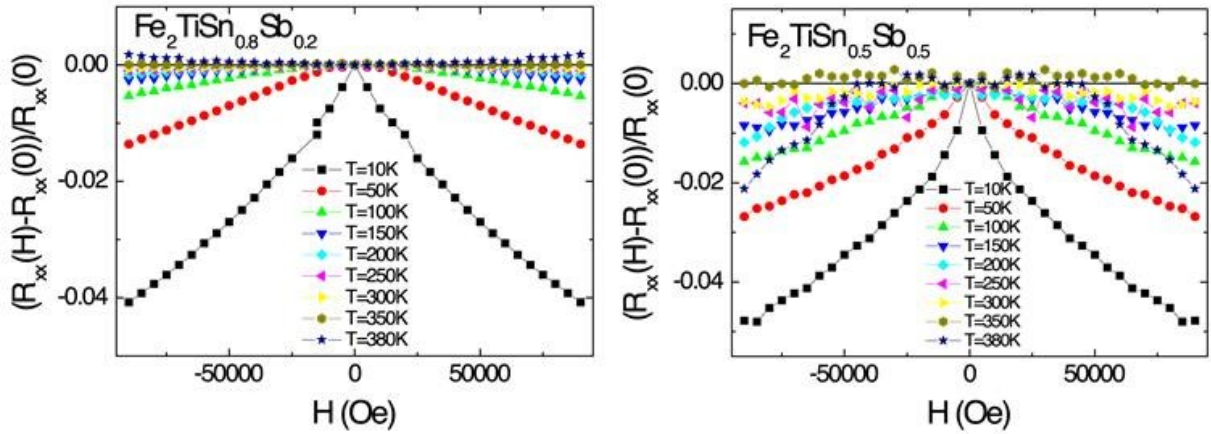

**Figure S4:** Longitudinal magnetoresistance curves versus applied field, measured in the x=0.2 and x=0.5 samples.

## 4. Technical details for the estimation of chemical potentials

Following FERE approach (ref. [18] in the main text), the reference chemical potentials $\mu_i^0$ were determined by fitting the experimental values (or theoretical values if experimental values were not available) of formation enthalpies for the binary and ternary compounds in Fe-V/Ti-Al/Sn(Sb) families of materials. The considered compounds were: $FeTi$, $Fe_2Ti$, $Fe_3Al$, $FeV_3$, $FeSb_2$, $TiAl$, $TiAl_3$, $Ti_6Sn_5$, $TiSb$, $Ti_3Sb$, $V_3Sn$, $VAl_3$, $Fe_2VAl$, $Fe_2TiAl$, $Fe_2VSn$, $Fe_2TiSn$, and $FeTiSb$. The deviations $\Delta\mu_i$ for a given compound are determined from the equilibrium conditions at different points from the region of phase stability (Figure 4 in main text). These conditions are imposed to allow the formation of $Fe_2TiSn$ and to prevent the formation of elemental bulk phases, binary phases, and ternary $FeTiSb$ and $FeTiSn$ hexagonal phases with space group $P6_3mc$. Under the assumption that $Fe_2TiSn$ is always stable, the elemental chemical potentials for Fe, Ti, Sn, and Sb are constrained by the following equilibrium equations:

Point A (Fe rich condition $\Delta\mu_{Fe} = 0$):

$2\mu_{Fe} + \mu_{Ti} + \mu_{Sn} = \mu_{Fe2TiSn}$; $2\mu_{Fe} + \mu_{Ti} \leq \mu_{Fe2Ti}$; $\mu_{Fe} = \mu_{Fe}^0 =>$

$\mu_{Ti} \leq \mu_{Ti}^0$; $\mu_{Sn} \geq \mu_{Sn}^0 - 2.42eV$; $\mu_{Sb} \leq \mu_{Sb}^0$;

Point B (Ti rich condition $\Delta\mu_{Ti} = 0$):

$2\mu_{Fe} + \mu_{Ti} + \mu_{Sn} = \mu_{Fe2TiSn}$; $\mu_{Fe} + \mu_{Ti} \leq \mu_{FeTi}$; $\mu_{Ti} = \mu_{Ti}^0 =>$

$\mu_{Fe} \leq \mu_{Fe}^0 - 0.28eV$; $\mu_{Sn} \geq \mu_{Sn}^0 - 1.69eV$; $\mu_{Sb} \leq \mu_{Sb}^0$;

Point C (Ti rich condition $\Delta\mu_{Ti} = 0$):

$2\mu_{Fe} + \mu_{Ti} + \mu_{Sn} = \mu_{Fe2TiSn}$; $6\mu_{Ti} + 5\mu_{Sn} \leq \mu_{Ti6Sn5}$; $\mu_{Ti} = \mu_{Ti}^0 =>$

$\mu_{Sn} \leq \mu_{Sn}^0 - 0.97eV$; $\mu_{Fe} \geq \mu_{Fe}^0 - 0.64eV$; $\mu_{Sb} \leq \mu_{Sb}^0$;

Point D (Sn rich condition $\Delta\mu_{Sn} = 0$):

$2\mu_{Fe} + \mu_{Ti} + \mu_{Sn} = \mu_{Fe2TiSn}$; $2\mu_{Ti} + 3\mu_{Sn} \leq \mu_{Ti2Sn3}$; $\mu_{Sn} = \mu_{Sn}^0 =>$

$\mu_{Ti} \leq \mu_{Ti}^0 - 0.96eV$; $\mu_{Fe} \geq \mu_{Fe}^0 - 0.65eV$; $\mu_{Sb} \leq \mu_{Sb}^0$;

Point E (Sn and Sb rich conditions $\Delta\mu_{Sn} = 0$, $\Delta\mu_{Sb} = 0$):

$2\mu_{Fe} + \mu_{Ti} + \mu_{Sn} = \mu_{Fe2TiSn}; \; \mu_{Fe} + \mu_{Sn} \leq \; \mu_{FeSn}; \; \mu_{Fe} + 2\mu_{Sb} \leq \; \mu_{FeSb2}; \; \mu_{Sn} = \mu_{Sn}{}^{0}; \; \mu_{Sb} = \mu_{Sb}{}^{0} =>$

$\mu_{Fe} \leq \; \mu_{Fe}{}^{0} - 0.54eV; \; \mu_{Ti} \geq \; \mu_{Ti}{}^{0} - 1.17eV$

Point F (Ti and Sn rich conditions $\Delta\mu_{Ti} = 0$, $\Delta\mu_{Sn} = 0$):

$2\mu_{Fe} + \mu_{Ti} + \mu_{Sn} = \mu_{Fe2TiSn}; \; \mu_{Fe} + \mu_{Ti} + \mu_{Sb} \leq \; \mu_{FeTiSb}; \; \mu_{Fe} + \mu_{Ti} + \mu_{Sn} \leq \; \mu_{FeTiSn}; \; \mu_{Ti} = \mu_{Ti}{}^{0}; \; \mu_{Sn} = \mu_{Sn}{}^{0} =>$

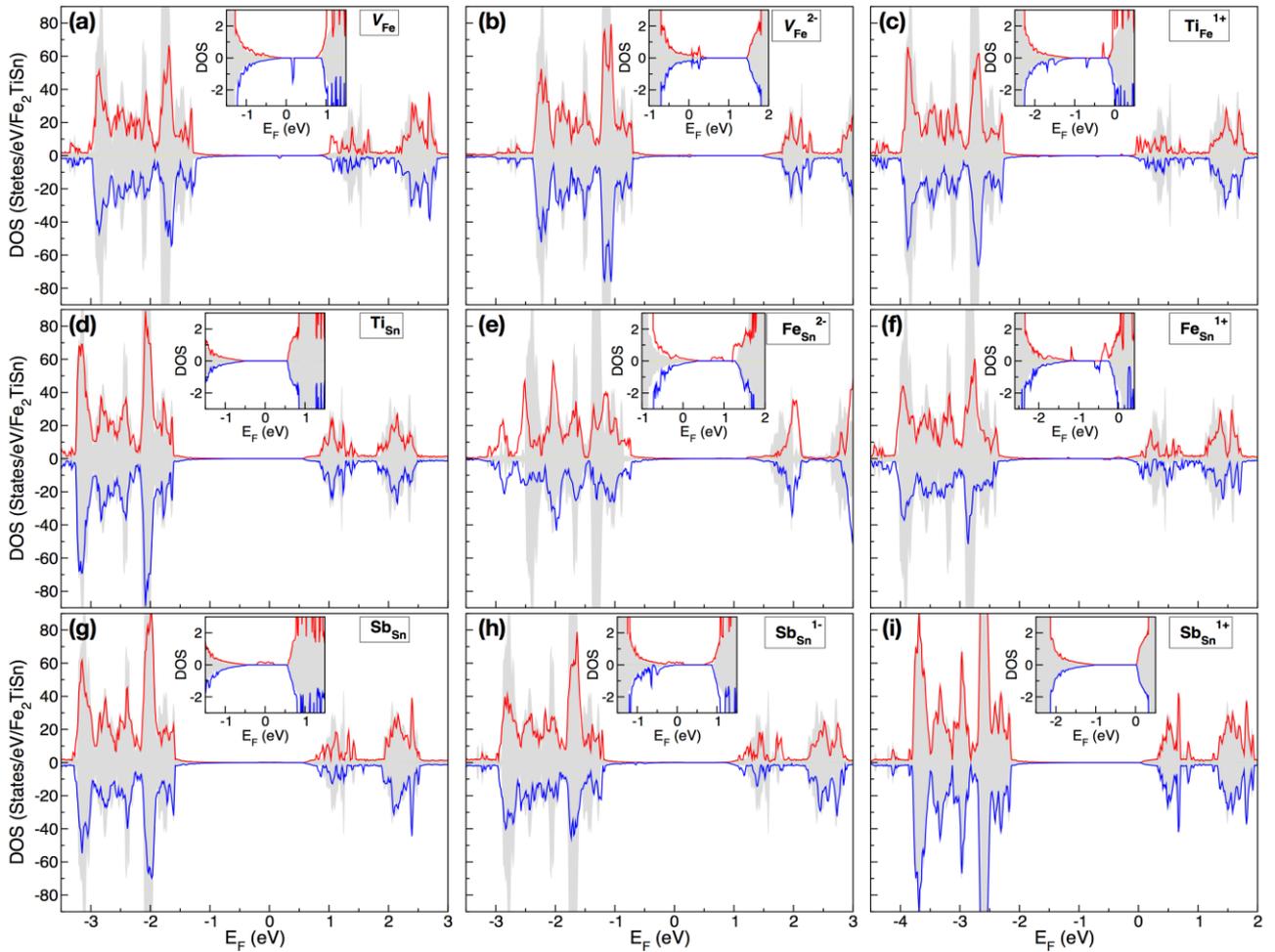

$\mu_{Fe} = \mu_{Fe}{}^{0} - 1.13eV; \; \mu_{Sb} \leq \; \mu_{Sb}{}^{0} - 0.08eV$

**Figure S5:** Electronic density of states (DOS) for spin up (red color) and spin down (blue color) channels of (a) $V_{Fe}$, (b) $V_{Fe}{}^{2-}$, (c) $Ti_{Fe}{}^{1+}$, (d) $Ti_{Sn}$, (e) $Fe_{Sn}{}^{2-}$, (f) $Fe_{Sn}{}^{1+}$, (g) $Sb_{Sn}$, (h) $Sb_{Sn}{}^{1-}$, and (i) $Sb_{Sn}{}^{1+}$. DOS of $Fe_2TiSn$ is shown in background gray colour. The electronic states near $Fe_2TiSn$ band gap are shown in the insets. $E_F = 0$ represents the energy position of the different defects.